\documentclass[12pt]{article}

\usepackage{graphicx}
\usepackage{indentfirst}
\usepackage{psfrag}

\begin{document}


\newpage

\thispagestyle{empty}

\begin{center}

{\bf \Large An overview of simply pulsed qubits}

\vskip 3em

{\ A. Chalastaras, L. Kaplan, Kh. Kh. Shakov, M. Smith, and J. H.
McGuire}

\vskip 1em

\textit{Physics Dept., Tulane University, New Orleans, LA 70118, USA}

\end{center}

\vskip 2em

\begin{center} \today \end{center}

\vskip 4em

{\bf Abstract}: The behavior of simply pulsed qubits (quantum
systems with two linearly independent states) may be characterized
by the energy difference $\Delta E$ between the two states of the
qubit and by an external stimulating potential $V(t)$ that causes
transitions between them.  Thus, the operation of such quantum
mechanical systems may be categorized in various regions that
explicitly depend on $\Delta E$ and $V(t)$. Limiting cases of
degenerate, perturbative, and adiabatic regions are discussed. A
comprehensive and illustrative map for simply pulsed qubits is
presented that can be used as a visual tool for students.
Furthermore, analytic solutions may be obtained when the interaction
$V(t)$ is proportional to $\delta(t-t_k)$, namely when a fast
interaction, called a kick, is used.

\newpage

\section{Introduction}

\indent Just as classical information and computation are built upon on-off bits, quantum information and
computation may be constructed on the basis of two-state (on-off) quantum systems known as
qubits~\cite{nielsen,waser}. These quantum building blocks are used to describe and to control atomic and
molecular reactions~\cite{malinovsky, smith}, electron dynamics~\cite{mcbook}, and time ordering in quantum
systems~\cite{progress}. However, properties of dynamic two-state quantum systems, such as transition rates from
one state to the other, are usually found numerically since analytic solutions are relatively rare. In this
paper we describe some simply pulsed qubits where the occupation probability of each of the two states, i.e.,
the likelihood of being in the on state or the off state, can be easily controlled.

\indent The difference between classical and quantum bits is that the classical occupation probability of the on
state must be either 1 or 0 (definitely on or definitely off) at \textit{all} times, while for a qubit the
probability $P(t)$ of being on may be any number between $0$ and $1$ until an observation is made.  That is,
before an observation the qubit may be in a combination or superposition of on and off states, similar to a
superposition of classical waves. After a qubit is observed, $P(t)$ collapses to 1 or 0 for that qubit. If $N$
qubits (where $N$ is large), initially all in the off state, are switched by the same external potential and
then observed at time $t$, $N \cdot P(t)$  of them will be found to be in the on state and $N \cdot (1-P(t))$
will remain in the off state.  It is this feature, namely that before an observation a qubit can be
simultaneously on \textit{and} off, that distinguishes the qubit from the classical bit that simply switches
between on or off (\textit{never} being in a superposition of on and off states). In this regard, the
mathematical rules (i.e., logic) differ for qubits and bits. The more sophisticated qubits can in principle
perform more complex operations~\cite{nielsen, roadmap}.

\section{Qubits}

\indent Since bits can only be on or off (e.g., corresponding to small magnets that can be magnetized in one of
two directions), they are simpler than qubits. Before the measurement, a qubit can be in both of its two states
at the same time. Reliably manipulating this coherent linear superposition of states, e.g., with external
fields, is essential for quantum computation~\cite{nielsen} and quantum control~\cite{malinovsky}, but often
difficult to achieve. The phenomenon of quantum parallelism~\cite{roadmap} allows a quantum processor to execute
actions in parallel (performing many classical computations at the same time). The classical computer, no matter
how fast, can only execute commands in series, which limits its computational power. Hence, a quantum computer
is more suitable for solving certain computational problems such as factorization of large numbers and database
search. Indeed, computer algorithms specifically designed to take advantage of the phenomenon of quantum
parallelism in qubits have been developed~\cite{shor, grover}. Various physical systems that are good candidates
for qubits are listed at the end of Section 2.

\subsection{Conceptual Description}

\indent Here we describe a conceptually simple way of controlling a qubit, that retains the key features of
manipulating more general qubits. The simplicity of the following description helps to clarify the concepts
without losing the essence of more complicated quantum systems.

    We consider a two-state system with possible energies $E_1$ and $E_2$,
where $\hbar\omega=\Delta E=E_2-E_1$ is the energy splitting between these two states. Since the zero point of
overall energy is arbitrary, the two eigenergies may be written as $+\Delta E/2$ and $-\Delta E/2$ without loss
of generality. The unperturbed Hamiltonian $\hat{H_0}$ does not couple the two eigenstates, i.e., $\hat{H_0}$
does not cause transition of population between the two states of the qubit. Such a system by itself is not very
interesting since the observable properties do not evolve with time. For example, if the system is initially
populated with any population number $N_1 = N \cdot P_1$ in state 1 and population $N_2 = N \cdot P_2$ in state
2, then without an external switching potential $N_1$ and $N_2$ do not change with time.

    What makes a qubit interesting and useful is the ability to change the populations of its
states in a controlled way.  This is desirable, for example, in a device such as a computer, where one wishes to
switch the state of a storage location back and forth from on to off at will. In classical computers, the logic
of these binary operations is simple, and is described in terms of Boolean algebra~\cite{boolean}, which is a
formal way of working with states that may be described as either on or off.  In such a classical device, the on
or off state is usually implemented physically by an on or off magnetic field at a given storage location.

\indent Quantum information is similar, except that the system is generally in a linear combination (linear
superposition) of on and off states. In both classical bits and qubits, the switching between on and off states
is done with an external field $V$. Since the state is to be switched from on to off at various times, $V$ is
some function of time $t$, i.e., $V = V(t)$. It is this external switching field $V(t)$ that enables a computer
to manipulate information, performing logic operations dynamically.  In our model we choose a $V(t)$ that simply
couples the two states without changing the eigenenergies $+ \Delta E/2$ and $- \Delta E/2$ of the unperturbed
Hamiltonian $\hat{H_0}$. While there are more complex ways to manipulate a two-level system, this is sufficient
to illustrate how a simple qubit works.

\subsection{Mathematical Description}

\indent In quantum mechanics, the state of a system at any time $t$ is fully described by its wave function
$\psi(t)$. If there are two linearly independent states of the system, labeled by $\left[ \begin{array}{c} 1 \\
0\end{array} \right]$ (e.g., corresponding to on) and $\left[\begin{array}{c} 0 \\ 1 \end{array} \right]$,
(e.g., corresponding to off), then in the most general case $\psi(t)$ is a linear superposition of these two
states. Mathematically this corresponds to

\begin{eqnarray}
\label{psi} \psi(t) = a_1(t) \left[
\begin{array}{c} 1 \\ 0 \end{array} \right] + a_2(t) \left[
\begin{array}{c} 0 \\ 1 \end{array} \right]
=\left[
\begin{array}{c} a_1(t) \\ a_2(t) \end{array} \right]
\,.
\end{eqnarray}

\noindent Here $a_1(t)$ and $a_2(t)$ are the probability (wave-like) amplitudes such that the probability of
finding $\psi(t)$ in the on state is $P_1(t) = |a_1(t)|^2$, and the probability that $\psi(t)$ is in the off
state is $P_2(t) = |a_2(t)|^2$. This is analogous to linear superpositions of electromagnetic fields in
classical wave mechanics, where the observed intensity is proportional to the square of either the electric or
the magnetic field. Normalization condition in quantum mechanics requires the constraint $P_1(t) + P_2(t)=1$
(conservation of probability), meaning that no dissipation\footnote{Dissipation corresponds to leakage of some
population out of the two-state system. While dissipation can be a significant problem in some practical
applications, in this paper we assume it is negligible.} has occurred since the system was initially formed.

A two-state system can be coupled (i.e., the particle population can
be transferred between the two states) by applying an external
potential $V(t)$. In this paper, the external potential has the
shape of a single pulse that can be sensibly characterized by a
single time duration, $\tau$.  A simple pulse could be for example a
Gaussian, an instantaneous kick (delta function), or a rectangular
pulse. The energy of the system is described by the full Hamiltonian
$\hat{H}(t)$, which includes the energy difference term $\Delta E$
and a time-dependent external potential term $V(t)$. Here we take
$\Delta E$ to be constant in time, corresponding to most
applications. Hence all the time dependence in the system's
Hamiltonian $\hat{H}(t)$ comes from $V(t)$. This full Hamiltonian
can be written in terms of the Pauli spin matrices (problem 1),
namely

\begin{eqnarray}
 \label{genh0} \hat H(t) &=& \hat H_0 + \hat V(t) \nonumber\\
           &=& \left[ \begin{array}{cc} -\Delta E/2 & 0 \\ 0 & +\Delta E/2
\end{array} \right] + \left[ \begin{array}{ccc} 0 & V(t) \\ V(t) &
0 \end{array} \right]  \\
 &=& -{\Delta E \over 2} \sigma_z + V(t) \sigma_x, \nonumber
\end{eqnarray}

\noindent where the widely used Pauli spin matrices~\cite{arfken} are defined as

\begin{equation}
\label{pauli}
 \\ I = \left[\begin{array}{rr} 1 & 0 \\ 0 & 1 \end{array} \right], \hspace{2pt}
 \sigma_x = \left[\begin{array}{rr} 0 & 1 \\ 1 & 0
 \end{array}
 \right],
 \hspace{2pt} \sigma_y = \left[\begin{array}{rr} 0 & -i \\ i
 & 0 \end{array} \right], \hspace{2pt} \sigma_z =
 \left[\begin{array}{rr} 1 & 0 \\ 0 & -1 \end{array} \right].
\end{equation}

\noindent Inserting the Hamiltonian $\hat{H}(t)$ from expression (\ref{genh0})
and the wave function $\psi(t)$ from expression (\ref{psi}) into the
time-dependent Schr\"{o}dinger equation,

\begin{equation} \label{schrod}
 \\ i\hbar\frac{d\psi(t)}{dt} = \hat{H}(t)\psi(t),
\end{equation}

\noindent we obtain (problem 2) a differential equation for the on and off probability amplitudes, $a_1(t)$ and
$a_2(t)$, as functions of the energy splitting and the external potential:

\begin{equation}
\label{eqmo} i\hbar \left [ \begin{array}{c} \dot a_1(t) \\ \dot
a_2(t)\end{array} \right] = \left[
\begin{array}{cc} -\Delta E/2 & V(t) \\ V(t) & +\Delta E/2 \end{array} \right ]
\left [ \begin{array}{c} a_1(t) \\ a_2(t)\end{array} \right] \,.
\end{equation}

\noindent Separating the equations, we can also write,

\begin{equation}
\label{eqmo1}\left.
\begin{array}{lll}
i\hbar\displaystyle\frac{da_1(t)}{dt}  =  i\hbar\ \dot a_1(t) &=& -\displaystyle\frac{\Delta E}{2} a_1(t)+ V(t)a_2(t)\\[10pt]
i\hbar\displaystyle\frac{da_2(t)}{dt} = i\hbar\ \dot a_2(t) &=& V(t)a_1(t) + \displaystyle\frac{\Delta E}{2} a_2(t)
\end{array}
\right\}.
\end{equation}

\newpage

\indent Conceptually and mathematically, it is convenient to isolate the time dependence of the system by
defining the evolution operator~\cite{sakurai} or matrix $\hat{U}(t, t_0)$ that connects the initial state
$\psi(t_0)$ with the final state $\psi(t)$,

\begin{equation}
\label{psievol}
\psi(t)=\hat{U}(t, t_0)\psi(t_0) \,.
\end{equation}

\noindent All the time dependence of the system is contained in the time evolution operator (or Green's
function) $\hat U(t,t_0)$, while the initial conditions are specified in $\psi(t_0)$.  Throughout the rest of
this paper, we take $t_0 = 0$ and write the evolution operator as $\hat{U}(t, t_0) = \hat{U}(t)$.  Then, the
evolution operator in equation (\ref{psievol}), has the following matrix representation

\begin{eqnarray}
\label{uoperatorzero} \hat{U}(t) = \left [
\begin{array}{cc}
  U_{11}(t,t_0)  & U_{12}(t,t_0)  \\
  U_{21}(t,t_0)  & U_{22}(t,t_0)
\end{array} \right] = \left [
\begin{array}{cc}
  U_{11}(t)  & U_{12}(t)  \\
  U_{21}(t)  & U_{22}(t)
\end{array} \right].
\end{eqnarray}

\noindent It is then straightforward to express the time-dependent probability amplitudes, $a_1(t)$ and
$a_2(t)$, using the evolution operator $\hat U(t)$, namely

\begin{eqnarray}
\label{amps} \left [ \begin{array}{c} a_1(t) \\ a_2(t) \end{array} \right]
 = \hat{U}(t) \left [ \begin{array}{c} a_1(0) \\ a_2(0) \end{array}
\right]\ = \left [ \begin{array}{cc}
  U_{11}(t)  & U_{12}(t)  \\
  U_{21}(t)  & U_{22}(t)
\end{array} \right]
\left [ \begin{array}{c} a_1(0) \\ a_2(0) \end{array} \right] \,.
\end{eqnarray}

\noindent The evolution operator $\hat U(t)$ may be obtained by solving a differential equation very similar to
(\ref{eqmo}):

\begin{equation}
\label{eqmou} i\hbar
 \left [ \begin{array}{cc} \dot U_{11}(t) &
\dot U_{12}(t) \\ \dot U_{21}(t) & \dot U_{22}(t) \end{array} \right] = \left[
\begin{array}{cc} -\Delta E/2 & V(t) \\ V(t) & +\Delta E/2 \end{array} \right ]
 \left [ \begin{array}{cc}  U_{11}(t) &
 U_{12}(t) \\  U_{21}(t) & U_{22}(t) \end{array} \right]
\,,
\end{equation}

\noindent and we require the time evolution operator to start out as the identity matrix $I$,
\begin{equation}
\hat{U}(0) =
 \left [ \begin{array}{cc}
  U_{11}(0)  & U_{12}(0)  \\
  U_{21}(0)  & U_{22}(0)
\end{array} \right] = I =
 \left [ \begin{array}{cc}
  1 & 0   \\
  0 & 1
\end{array} \right]
\end{equation}
so that $\psi(t)=\psi(0)$ at the initial time $t=t_0=0$.

\noindent Formally, the solution to equation (\ref{eqmou}) is given by
\begin{eqnarray}
\label{uoperator} \hat{U}(t) = \left [ \begin{array}{cc}
  U_{11}(t)  & U_{12}(t)  \\
  U_{21}(t)  & U_{22}(t)
\end{array} \right] = Te^{-i \int_{0}^{t} \hat{H}(t') dt' / \hbar} = Te^{-i \int_{0 }^{t} [\hat H_0+ \hat{V}(t')] dt' / \hbar} \,,
\end{eqnarray}

\noindent where $T$ is the Dyson time ordering symbol~\cite{goldwatson,merzbacher} that enforces $\hat{H}(t_1)$
acting on the system prior to $\hat{H}(t_2)$ if $t_1<t_2$.

Equations (\ref{psi}), (\ref{genh0}), (\ref{schrod}), (\ref{psievol}), (\ref{amps}), (\ref{eqmou}), and
(\ref{uoperator}) form the basic equations for a qubit.  These equations are governed by the energy splitting
$\Delta E$ and the external potential $V(t)$. Depending on the complexity of the time evolution operator, these
equations may or may not have analytic solutions. In the latter case, numerical solutions of the coupled
differential equations (\ref{eqmo1}) may obscure information about these quantum systems. Analytic solutions,
when possible, are more convenient and easy to analyze.

\subsection{Practical Considerations for a System of Qubits}

\indent Up to this point, we have treated the qubit as an abstract mathematical structure. However, in
applications some practical issues have to be addressed. For a system of qubits to be capable of performing
quantum computation~\cite{nielsen,progress,roadmap}, control of population transfer between the two states of
the qubit is not sufficient.  The following list, dubbed the ``DiVincenzo checklist"~\cite{divcenzo} after its
developer in 1997, addresses some practical issues for a quantum computer:

\begin{itemize}
    \item A large number of \textit{coupled} qubits need to be
reliably controlled.

\item It should be possible to prepare the qubits in a desired initial state.
Commonly, in the initial state all qubits are on or all qubits are off.

    \item Decoherence should be minimized. In terms of the energy splitting
$\Delta E$ between the two states, the decoherence time $t_{d}$ over which quantum phase information is lost
must satisfy $t_{d}\gg 2\pi\hbar/\Delta E$~\cite{waser}.

    \item Quantum gates need to be designed that will control the operation of
the qubits (these gates play a role similar to classical gates).

    \item The information contained in the qubits must be extractable at the
end of the computation if the outcome of the qubit operations is to be used in
a productive manner.
\end{itemize}

A number of quantum systems satisfy these requirements. Optical
photons~\cite{imayama}, nuclear spins~\cite{divincenzo2, cory, gershenfeld},
ion traps~\cite{cirac}, nuclear magnetic resonance (NMR)~\cite{schulman}, and
electrons on superfluid helium \cite{platzman} are some commonly discussed
examples.

\indent The \textit{QIST Quantum Computation Roadmap}~\cite{roadmap} addresses
the key issues in quantum computation and the latest advances in the field.  In
the last few years, considerable progress has been made in overcoming the
problem of decoherence, i.e., loss of phase control. Decoherence is an
important but complicated issue and is discussed elsewhere~\cite{rau}.

\indent Next, we discuss different physical limits for simply pulsed qubits and the graphical qubit map is
introduced.

\section{Map}

\subsection{Significance of the Qubit Map}

\indent A qubit map, such as that shown in Figure 1, is a tool that enables one to visualize how the behavior of
simply pulsed qubits may depend on the variables $\Delta E$ and $V(t)$.  In order to make the qubit map a
helpful visual tool, the map coordinates are taken to be ${\Delta E\tau}/{2\hbar}$ (where $\tau$ is the time
duration of $V(t)$) and $\int^\infty_0 V(t')dt'/\hbar$. These two variables are dimensionless and range in
absolute value from 0 to $\infty$. The variable ${\Delta E\tau}/{2\hbar}$ determines the effect of the
unperturbed Hamiltonian $\hat{H_0}$ on the evolution operator $\hat{U}(t)$, while $\int^\infty_0 V(t')dt'/\hbar$
measures the influence of the external potential on $\hat{U}(t)$.  It is useful to think of these variables as
independent phase angles or action-like integrals \cite{messiah,goldstein}.

    It should be noted here that the boundaries separating the qubit map
regions are not precisely defined. The regions overlap each other (see curly lines in the map) when either of
the two phases ${\Delta E\tau}/{2\hbar}$ or $\int^\infty_0 V(t')dt'/ \hbar$ is of the order of $2\pi$. This
behavior should not be surprising since the definitions of the degenerate, perturbative, and adiabatic limits
are of an approximate nature. The map helps to identify the regions where qubit behavior may be described
analytically, and indicate ballpark values of the parameters for which various analytic solutions are
applicable.  For instance, if we wish to investigate slow weakly perturbed qubits, we look for large phases
$\Delta E\tau/{2\hbar}$ and small $\int^\infty_0 V(t')dt'/\hbar$. For strongly perturbed degenerate qubits, we
look for small $\Delta E\tau/{2\hbar}$ but large values of $\int^\infty_0 V(t')dt'/\hbar$.

The arrow pointing toward the lower left corner of the Figure indicates the ``fast" or short-time limit $\tau
\to 0$. Here the time $\tau$ of the pulse is too short to allow for significant phase accumulation  due either
to the external potential $V(t)$ or to the energy splitting $\Delta E$, i.e., both $\Delta E\tau/{2\hbar}$ and
$\int^\infty_0 V(t')dt'/\hbar$ are small. The opposite $\tau \to \infty$ or ``slow" limit indicated by the arrow
pointing toward the upper right corner is where the pulse is long enough for both the energy splitting and the
external potential to have a large effect over the duration of the pulse.

\begin{figure}[p]
\begin{center}
\psfrag{xy}{$0$}
 \psfrag{x1}{$2 \pi$}
 \psfrag{x2}{$\infty$}
 \psfrag{x3}[c][c][1.5][0]{$\frac{\int_0^\infty {V(t')}
 dt'}{\hbar}$}
 \psfrag{y1}[c][c][1][90]{$2 \pi$}
 \psfrag{y2}[c][c][1][0]{$\infty$}
 \psfrag{y3}[c][c][1.5][90]{$\frac{\Delta E \tau}{2
 \hbar}$}
 \psfrag{z1}[l][l][1][0]{$\tau \rightarrow 0$ Fast}
 \psfrag{z2}[c][c][1][0]{$\tau \rightarrow \infty$ Slow}
\vspace*{-0.2in}
\includegraphics[keepaspectratio=true,width=1.0\textwidth]{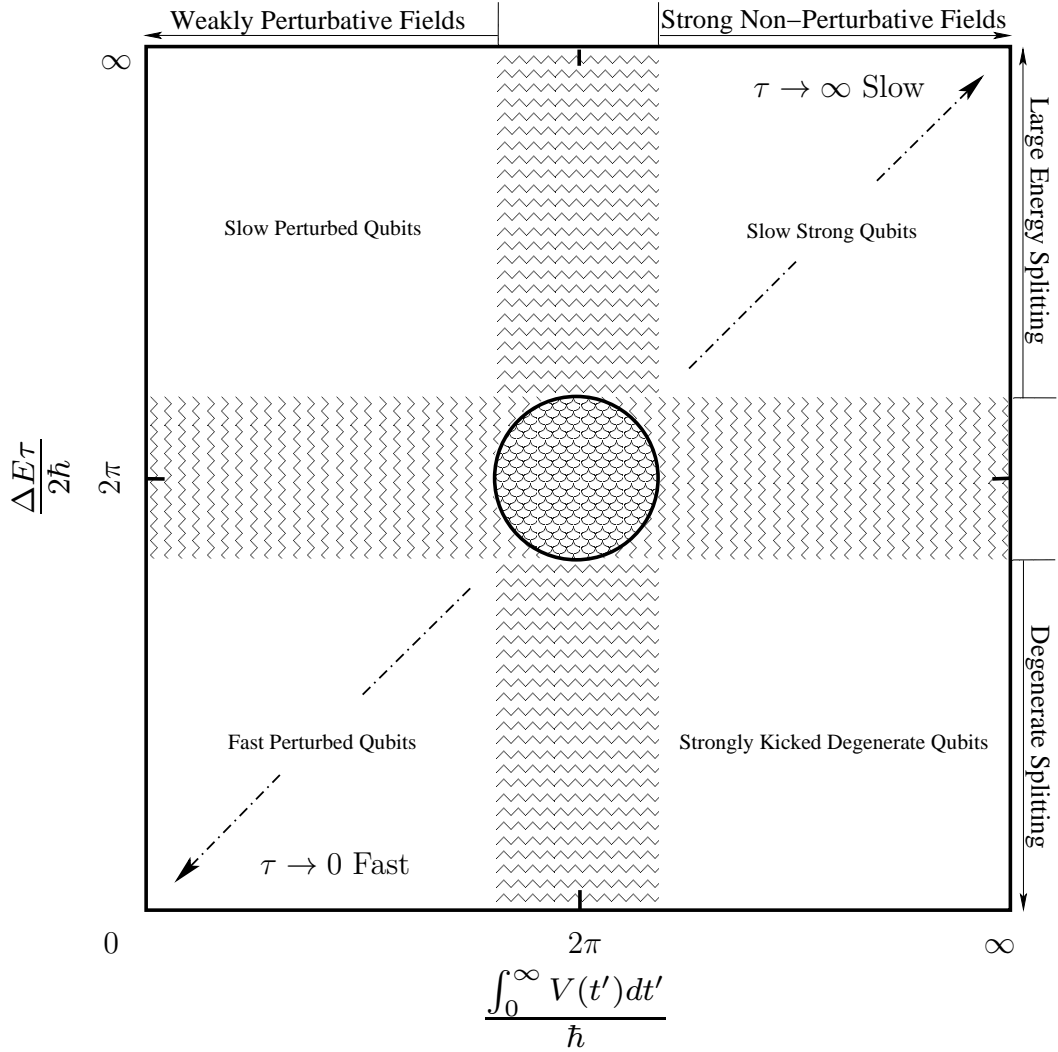}
\vspace*{-0.2in} \caption{\footnotesize{Qubit map for qubits
interacting with simply pulsed finite external potentials. Here
$\tau$ is the duration of the pulse, $\Delta E$ is the energy
difference between the two states of the qubit, and $V(t)$ is the
external potential. Note that the single pulse $V(t)$ is neither
harmonic nor periodic. The two axes are dimensionless. These axes
may be thought of as action-like~\cite{messiah,goldstein} phase
accumulations due respectively to the energy difference $\Delta E$
between the two possible qubit states and to the external potential
$V(t)$. Everywhere outside the central region, approximate analytic
solutions may be obtained.  When the total phase associated with the
external potential is small, i.e. $\int_0^\infty {V(t')} dt'/\hbar
\ll 2\pi$, then the expression (\ref{uoperator}) for the time
evolution operator may be expanded in powers of $V$ using
$e^{-iV(t')dt'/\hbar} \approx 1-iV(t')dt'/\hbar$ and only the first
few terms retained.  This corresponds to standard perturbation
theory for either quickly perturbed qubits where $\tau$ is small or
slowly perturbed qubits where $\tau$ is large. Similarly, if the
phase associated with the energy splitting is small, $\Delta
E\tau/2\hbar \ll 2\pi$, then we can treat the qubit as degenerate,
and the solution can be expanded in powers of $\Delta E$.  If both
phases are large, then the adiabatic approximation generally
applies.}} \end{center}
\end{figure}

\subsection{Regions of the Map}

\indent It is sensible to separate our map into four regions including the interesting central region and to
examine each one individually. Analytic solutions\footnote{While solutions of two-state systems with rapidly
changing external potentials were already examined about 40 years ago~\cite{smith}, the focus has now shifted to
quantum control.} exist only for some regions of the map. It should be emphasized that this map is valid only
for singly pulsed qubits. Multiply pulsed qubits~\cite{kaplantime}, including periodically pulsed qubits
\cite{eberly}, are not discussed here since an additional time parameter, e.g., the typical time between pulses
or the period of oscillation of the potential, would require a more complicated map.

\subsubsection{Degenerate Region}
\label{secdegen}

\indent When the energy splitting between the two states of the qubits is much less than the strength of the
external interaction (i.e., $\Delta E\ll V(t)$), we are in the degenerate region of the map.  The diagonal
extending from the lower left (``fast'') corner to the upper right (``slow") corner of the Figure forms the
boundary of the degenerate region: the behavior of the system is dominated by the external interaction $V(t)$
whenever we are to the right of or below this diagonal line.  The evolution of the system is then given by
simple sine and cosine functions, and the mathematical complexity is greatly reduced. It can easily be shown
(problems 3 and 4) that as $\Delta E\to 0$ in equation (\ref{uoperator}), the evolution matrix of the degenerate
qubit or dit\footnote{Our usage of the term ``dit" is not universal.}~\cite{shakovMc} becomes
\begin{equation}
\label{degen}
    \hat{U}(t) \to \hat{U}_{\rm dit}(t) =
 \left[\begin{array}{cc}
\cos \alpha &
-i \sin \alpha \\
-i \sin \alpha & \cos \alpha
\end{array} \right]  \ \ ,
\end{equation}
where
\begin{equation}
\alpha = {\int_0^t V(t') dt' \over \hbar}
\end{equation}

\noindent is the action integral. When $\alpha$ is an integer
multiple of $\pi/2$, the degenerate bit becomes a classical bit
where the occupation probabilities simply get switched between 1 and
0.

The analytic expression (\ref{degen}) is only valid when two conditions $\Delta E \tau/2\hbar  \ll \int_0^t
V(t') dt'/\hbar$ and $(\Delta E \tau/2\hbar)^2 \ll \int_0^t V(t') dt'/\hbar$ are simultaneously satisfied.  The
need for the second condition may be removed, thereby extending the solution to the entire degenerate region to
the right of the diagonal, by replacing $\alpha$ with
\begin{equation}
\theta = {\int_0^t \Omega(t') dt' \over \hbar} \,,
\end{equation}
where $\Omega(t')=\sqrt{V^2(t')+(\Delta E/2)^2}$. We notice that
$\Omega(t')=V(t')$ and thus $\theta=\alpha$ for $\Delta E \to 0$.

We will see below that the degenerate region overlaps other regions of the map, such as the perturbative,
adiabatic, and kicked regions. In these areas of overlap, two or more analytic solutions become mathematically
equivalent.

\subsubsection{Perturbative Region}

When the external potential $V(t)$ is small, we are in the perturbative region. The perturbative region includes
both upper and lower left quadrants of the qubit map. In this particular region $\int_0^\infty V(t') dt'/{\hbar}
\ll 2\pi$.  The external interaction $V(t)$ is sufficiently weak that the qubit remains essentially in its
initial on or off state, acquiring only an overall phase associated with the energy $\pm \Delta E/2$.  It can be
shown (problem 5a) that the evolution operator in equation (\ref{uoperator}) becomes
\begin{equation}
\label{perturb}
\!\!\!\!    \hat{U}_{\rm pert}(t) =
 \left[\begin{array}{cc}
e^{it(\Delta E /2\hbar)} &
-i\int_0^\infty e^{\frac{i(t-2t')\Delta E}{2\hbar}}V(t')dt'/\hbar \\
-i\int_0^\infty e^{\frac{-i(t-2t')\Delta E}{2\hbar}}V(t')dt'/\hbar & e^{-it (\Delta E /2\hbar)} \end{array}
\right]
\\.
\end{equation}

\noindent For the extreme special case where $V(t)= 0$ at all times(problem 5b), the evolution operator in
equation (\ref{perturb}) becomes

\begin{equation}
\label{zero}
    \hat{U}(t) \to \hat{U}_{\rm zero}(t) =
 \left[\begin{array}{cc}
 e^{it (\Delta E /2\hbar)}&
0 \\ 0 & e^{-it (\Delta E/2\hbar)}
\end{array} \right]  \\.
\end{equation}

\noindent Setting $V(t)= 0$ uncouples the differential equations in expression
(\ref{eqmo1}) making the math trivial. We obtain $a_1(t)=a_1(0) e^{it\Delta
E/2\hbar}$ and $a_2(t)= a_2(0) e^{-it \Delta E/2\hbar}$, i.e. the amplitudes of
being in the on or off state acquire phases as time evolves but the
probabilities $|a_1(t)|^2$ and $|a_2(t)|^2$ remain what they were at the
initial time $t=0$.

\indent The perturbative region is of little interest in our case since the
system does not change appreciably from its original state. In the case of
zero external potential, the qubit does not change its state at all.

\subsubsection{Adiabatic Region}

The adiabatic region covers the upper right, upper left, and lower right quadrants of the qubit map, excluding
only part of the lower left (``fast") quadrant and the central region. The explicit condition for adiabaticity
is~\cite{kaplantime}
\begin{equation}
\label{adiabatic} \hbar \dot{V}(t)\Delta E \ll [V^2(t)+ (\Delta E/2)^2]^{3/2} \,,
\end{equation}
which can be further simplified in the degenerate and nondegenerate limits. In the degenerate adiabatic case
$\Delta E \ll V(t)$, so (\ref{adiabatic}) yields the condition $\hbar \dot{V}(t)\Delta E \ll V^3(t)$ or $\hbar
\Delta E /\tau \ll V^2(t)$. The time derivative of the external potential is of the order $V(t)/\tau$, i.e.,
$\dot V \sim V/\tau$, which is the typical slope of the single-pulse potential. In the nondegenerate adiabatic
case $\Delta E \gg V(t)$, and the validity condition (\ref{adiabatic}) becomes $\hbar \dot{V}(t) \ll (\Delta
E)^2$ or $\hbar V(t)/\tau \ll (\Delta E)^2$.

 \indent The meaning of the adiabatic region is that the external potential
changes slowly, so that the quantum system is able continuously to adjust to the new Hamiltonian.  Consider for
example a classical system of two pendulums connected by a spring.  The system can oscillate in one of two
normal modes: a lower-frequency swinging mode where the two pendulums move left and then right in unison, and a
higher-frequency vibrational mode where the spring is alternately stretched and compressed.  If we move the
pivot point of one or both pendulums slowly and carefully enough, the pendulums continue to oscillate in the
same (lower or higher) mode they started in, because not enough energy is provided by the external perturbation
to switch the motion from the lower-frequency oscillation to the higher-frequency oscillation or vice versa.
Similarly, an adiabatically driven qubit that is initialized in the lower energy ($-\Delta E/2$ or off) state
before the pulse is turned on will at any future time remain in the lower energy state of the full Hamiltonian
$\hat H(t)$. After the pulse is over and the Hamiltonian returns to the unperturbed Hamiltonian $\hat H_0$, the
qubit must return to the off state with energy $-\Delta E/2$.  Thus, the adiabatic region is of no great
interest in the case of pulsed qubits since we wish to completely transfer the population of a qubit from one
state to the other.

Mathematically, the evolution operator for an adiabatically driven qubit is
\begin{equation}
\label{uadiab}
     \hat{U}_{\rm adiab}(t) =
\left[\begin{array}{cc}
\cos \theta(t) + i {\Delta E\over 2\Omega(t)} \sin \theta(t) &
-i {V(t) \over \Omega(t)} \sin \theta(t) \\
-i {V(t) \over \Omega (t)} \sin \theta(t) &
\cos \theta(t) - i {\Delta E\over 2\Omega(t)} \sin \theta(t)
\end{array} \right]\,,
\end{equation}
where $\theta(t)$ and $\Omega(t)$ are defined as in Section~\ref{secdegen}, and for simplicity we assume
$V(t)=V(0)$, i.e., the external potential returns to its initial value at the end of the pulse.
Equation~(\ref{uadiab}) can be reduced (problem 6) to equations we presented above for the degenerate and zero
external potential cases.

\subsubsection{Central Region}
None of the limits discussed previously overlap with the central region of the map. Here the two independent
variables $\Delta E\tau/{2\hbar}$ and $\int^\infty_0 V(t') dt'/{\hbar}$ are both of order $2\pi$, and there is
no large or small parameter to simplify the solution of the problem. If the pulse has a particularly simple
(e.g., rectangular) shape, an analytic expression is possible~\cite{kaplantime} even in this region, but for a
general pulse shape no analytic solution exists and the evolution must be computed numerically.

\section{Kicked Qubit Approximation}

There is a useful approximation for simply pulsed qubits, namely the sharp, narrow pulse or ``kick" limit in
which the width $\tau$ of the pulse goes to zero, while the integrated strength or the area under the external
potential curve

\begin{equation}
\alpha_k = \int_0^t V(t') dt'/\hbar
\end{equation}

\noindent remains fixed. Formally, the shape of a very narrow pulse of finite total strength $\alpha$ may be
expressed by a delta function: $V(t')=\alpha_k \hbar\delta(t-t_k)$, where $t_k$ is the time at which the pulse
is applied. The kicked region corresponds to the lower half of the qubit map in Figure 1. Here the duration of
the pulse is so short that $\Delta E \tau/2\hbar \ll 2\pi$, i.e., there is not enough time for the energy
splitting $\Delta E$ to have a significant effect while the pulse is active. The integrated strength of the
pulse, $\alpha_k$, may be either large or small in this region.  If $\alpha_k$ is large, we are in the lower
right quadrant of the map, where the kicked region overlaps with the adiabatic region. If $\alpha_k$ is small,
we are in the lower left quadrant, where the kicked region overlaps with the perturbative region.

In the kicked region, the evolution of the qubit can be described as follows: (i) before the kick, the qubit
evolves in accordance with the unperturbed Hamiltonian $\hat H_0$, i.e. the on and off probability amplitudes
$a_1(t)$ and $a_2(t)$ acquire phases proportional to the energy splitting $\Delta E$ (as in (\ref{zero})), while
the on and off probabilities, given by $|a_1(t)|^2$ and $|a_2(t)|^2$, remain unchanged. (ii) For the duration of
the kick, the energy splitting associated with the unperturbed Hamiltonian $\hat H_0$ may be ignored, and a
transfer of population between on and off states may occur, depending only on the integrated strength $\alpha_k$
of the kick. (iii) Finally, free evolution governed by $\hat H_0$ resumes after the kick is complete, and no
further population transfer occurs. This sequence of events is very similar to the collision approximation
studied in introductory physics, where (i) a particle initially moves freely with constant velocity, (ii) then
undergoes an instantaneous collision with a wall, during which the velocity is changed but there is not enough
time for the particle to move, and (iii) finally, the particle once again resumes free flight with a (new)
constant velocity.

For such kicks, the integration over time in equation (\ref{uoperator}) is straightforward (problem 7) and the
time evolution matrix becomes a product of three parts,
\begin{eqnarray}
\hat{U}_{\rm kicked}(t) &=& e^{i{\Delta E\over 2\hbar}(t-t_k)\sigma_z} e^{-i
\int_{t_k - \epsilon}^{t_k + \epsilon} V(t') dt' \sigma_x / \hbar} e^{i{\Delta
E\over 2\hbar}t_k\sigma_z} \nonumber
\\
\label{kickedut} &=& \left [ \begin{array}{cc}
e^{i\Delta Et/2\hbar} \cos \alpha_k & -i e^{i\Delta E(t-2t_k)/2\hbar} \sin
\alpha_k \\ -i e^{-i\Delta E(t-2t_k)/2\hbar} \sin \alpha_k & e^{-i\Delta
Et/2\hbar} \cos \alpha_k
\end{array} \right ] \,.
\end{eqnarray}

\noindent Here, we may set\footnote{For a single kick setting $T\rightarrow1$ corresponds to the limit of no
time ordering, which is beyond the scope of this paper. For some useful discussions on time ordering see
reference~\cite{progress}.} the Dyson time ordering symbol $T\rightarrow1$ for the duration of the pulse
($t_k-\epsilon < t<t_k+\epsilon$).  It can also be shown (problem 8) that equation~(\ref{kickedut}) will reduce
to other limiting cases discussed previously.

\noindent Moreover, the occupation probabilities for a kicked qubit initially in state 1 are, from equation
(\ref{amps}),

\begin{eqnarray}
\label{Pk1}
   P_1(t) &=& |a_1(t)|^2 = |U_{11 \ \rm kicked}(t)|^2
    = \cos^2 \alpha_k  \nonumber \\
   P_2(t) &=& |a_2(t)|^2 = |U_{21 \ \rm kicked}(t)|^2
    = \sin^2 \alpha_k \ .
\end{eqnarray}

\noindent Since $\sin^2 \alpha_k \ + \cos^2 \alpha_k \ = 1$ for this
closed two-state system, the conservation of population holds, i.e.,
there is no dissipation. We notice also that the transfer
probability $P_2(t)$ does not depend on the energy splitting $\Delta
E$.  This simple instructional example may be extended to a series
of kicks~\cite{kaplantime}.  It is one of the few cases in which
analytic solutions may be obtained for qubits controlled by external
potentials.

\section{Summary}

\indent A primary motivation of this paper has been to give an overview of simply pulsed qubits. We have
developed a simple, yet comprehensive and instructive, two-dimensional map for such qubits. Analytic solutions
with limited applicability are found and their corresponding regions on the map are identified.  One promising
class of analytic solutions for dynamic qubits are the kicked solutions, valid for qubits subject to fast
interactions. These provide clear and useful examples to students and offer an alternative to the
well-established rotating wave approximation (RWA) \cite{eberly, shore, milonni}. The RWA method assumes a
sinusoidal external perturbation with frequency chosen to be in resonance with the desired transition between
two states.  The solutions presented in this paper are, in principle, valid for a wide spectrum of pulse widths
$\tau$ and energy splittings $\Delta E$.

\indent Qubits are building blocks for quantum computation and quantum information.  Their behavior and
interaction with each other and the environment have to be addressed before a quantum computer becomes a
reality. Ultimately, N-qubit systems and their interactions will have to be successfully controlled. The simply
pulsed qubits discussed in this paper may form a basic building block for more complex interconnected N-qubit
systems.

\newpage

\section{\textit{APPENDIX}\\{Problems for Students}}

\begin{itemize}

    \item \textbf{Problem 1}: (a) Show that a 2x2 matrix that corresponds to a two-state quantum mechanical
observable can be written in terms of the Pauli matrices. [Hint: Write down a generic Hermitian 2x2 matrix and
try to expand it as a sum of the Pauli matrices with appropriate prefactors]. (b) If the qubit is implemented as
a spin-1/2 particle, $\left[ \begin{array}{c} 1 \\ 0
\end{array} \right]$ represents a state where the spin points in the $+z$ direction, and $\left[
\begin{array}{c} 0 \\ 1 \end{array} \right]$ represents a state where the spin points in the $-z$ direction,
then what do operators $\sigma_x$, $\sigma_y$, and $\sigma_z$ correspond to physically?

    \item \textbf{Problem 2}: Using equations (\ref{psi}), (\ref{genh0}), and (\ref{schrod}), verify equation
(\ref{eqmo}).  Rules for simple matrix operations are available in many elementary math books and on the web.

    \item \textbf{Problem 3}: (a) Solve the evolution of the degenerate qubit
or dit by plugging $\Delta E = 0$ into equation (\ref{eqmo}), thus proving that the probability amplitudes,
$a_1(t)$ and $a_2(t)$, oscillate sinusoidally with $\alpha=\int_0^t V(t')dt'/\hbar$. (b) Make plots of the
probabilities $P_1=|a_1(t)|^2$ and $P_2=|a_2(t)|^2$. What can you deduce from the plots about dissipation in the
system? (c) Find the condition for complete transfer from state 1 to state 2.

    \item \textbf{Problem 4}: (a) Solve the evolution of the degenerate qubit or dit
by using the evolution operator in equation (\ref{uoperator}) and then plugging $\hat{U}(t)$ into equation
(\ref{amps}). [Hint: Since $\Delta E=0$, $\hat H_0$ vanishes and time ordering effects disappear ($T=1$). Then
$\hat U(t) = e^{-i\int_0^t V(t)\sigma_x/\hbar}$, and the following identities may be useful:
$e^{i\theta\sigma_z} = I \cos(\theta) + i\sigma_z \sin(\theta)= \left[\begin{array}{rr} e^{i\theta} & 0 \\ 0 &
e^{-i\theta}\end{array} \right]$ and $e^{i\theta\sigma_x} = I \cos(\theta) + i\sigma_x \sin(\theta)$.] (b) Do
you get the same probability amplitudes as in problem (3b)? (c) Check to see that the evolution operator is
unitary, that is $\hat U(t)^\dag \hat U(t)= I$.

    \item \textbf{Problem 5}: (a) Show that in the perturbative limit, the
evolution operator in equation (\ref{perturb}) satisfies equation (\ref{eqmou}). (b) Do the same for the case of
zero external potential. That is, prove equation (\ref{zero}). You may use the result from part (a).

 \item \textbf{Problem 6}: (a) Show that the adiabatic equation (\ref{uadiab}) for the evolution operator reduces to the degenerate
 equation (\ref{degen}) in the limit $\Delta E \to 0$. (b) Show that the adiabatic equation (\ref{uadiab}) reduces to the
zero potential equation (\ref{zero}) in the limit $V(t) \to 0$.

 \item \textbf{Problem 7}: Prove equation (\ref{kickedut}) by integrating and then multiplying out the exponentials in the first line, thus obtaining the
matrix in the second line of equation (\ref{kickedut}) [Hint: You may find useful the identities given in
problem (4)].

 \item \textbf{Problem 8}: (a) Show that equation (\ref{kickedut}) for a kicked qubit is equivalent to the perturbative
equation (\ref{perturb}) when the kick strength is small, i.e., $\alpha_k \ll 1$ [Hint: When $\alpha_k \ll 1$,
$\cos\alpha_k \approx 1$ and $\sin \alpha_k \approx \alpha_k $. In equation~(\ref{perturb}), replace $t'$ with
$t_k$ since we are in the kicked region in this problem]. (b) Show that the equation (\ref{kickedut}) for a
kicked qubit reduces to the degenerate equation (\ref{degen}) in the limit $\Delta E \to 0$.

\end{itemize}

\end{document}